\documentclass[prb,twocolumn,showpacs]{revtex4}

\usepackage{graphicx}
\usepackage{dcolumn}
\usepackage{bm}

\begin{document}


\title{Spectral properties near the Mott transition in the two-dimensional Hubbard model with next-nearest-neighbor hopping}

\author{Masanori Kohno}
\affiliation{International Center for Materials Nanoarchitectonics (WPI-MANA), 
National Institute for Materials Science, Tsukuba 305-0044, Japan}

\date{\today}

\begin{abstract}
The single-particle spectral properties near the Mott transition in the two-dimensional Hubbard model with next-nearest-neighbor hopping are investigated by using cluster perturbation theory. 
Complicated spectral features of this model are simply interpreted, by considering how the next-nearest-neighbor hopping shifts the spectral weights of the two-dimensional Hubbard model. 
Various anomalous features observed in hole-doped and electron-doped cuprate high-temperature superconductors are explained in a unified manner 
as properties near the Mott transition in a two-dimensional system whose spectral weights are shifted by next-nearest-neighbor hopping. 
\end{abstract}


\maketitle
\section{Introduction} 
Since the discovery of cuprate high-temperature superconductors, \cite{highTc} 
electronic properties of a two-dimensional (2D) system near the Mott transition have attracted considerable attention, 
because the high-temperature superconductors are obtained by doping layered-structure Mott insulators with holes or electrons. \cite{highTc,TokuraEleDope} 
The anomalously high superconducting-transition temperature ($T_c$) is considered to be related to the anomalous features near the Mott transition in a 2D system. \cite{AndersonRVB,DagottoRMP,NagaosaRMP} 
Because both hole-doped and electron-doped systems exhibit superconductivity, \cite{highTc,TokuraEleDope} it is natural to consider that the mechanism of high-$T_c$ superconductivity is included in both cases. 
Thus, a comprehensive understanding of the electronic properties of hole-doped and electron-doped systems is desired in order to elucidate the mechanism of high-$T_c$ superconductivity. 
\par
A remarkable difference in spectral feature between hole-doped and electron-doped systems near the Mott transition is the location of the pseudogap in the momentum space: 
the spectral weights around the Fermi level are considerably reduced around $(\pm\pi,0)$ and $(0,\pm\pi)$ in hole-doped systems, \cite{ShenRMP,FlatbandBi2212,LSCO_FS,Ca2NaCuO2Cl2} 
whereas those around $(\pm\pi/2,\pm\pi/2)$ are almost lost in electron-doped systems. \cite{ShenRMP,ArmitageRMP,ArmitagePRL} 
Although this difference has been more or less reproduced in the 2D Hubbard and $t$-$J$ models with next-nearest-neighbor hopping, the interpretations are controversial. \cite{KuskoMF,KusunoseWeakU,ModeCouping,KyungSuperAF,Kyung2PSC,KyungCDMFT,eledopeCPT,eledopeVCPT,CivelliCDMFT,AichhornVCA,eledopeDCA,eledopeFLEX,eledopeSBtJ,eledopeSBHub,tJTohyamaMaekawa,tJTohyama,tJPrelovsek,SakaiImadaPRB,DahnkenVCPT,FS_DCA,YanasePG} 
For the pseudogap around $(\pm\pi/2,\pm\pi/2)$ in electron-doped systems, the most widely accepted interpretation would be that it can be identified as the gap due to the band splitting 
caused by antiferromagnetic long-range order. \cite{KuskoMF,KusunoseWeakU,ModeCouping,AichhornVCA,eledopeDCA,eledopeFLEX,eledopeSBtJ,eledopeSBHub,tJTohyamaMaekawa,tJTohyama,tJPrelovsek} 
There are also other interpretations, such as that it is the result of short-range antiferromagnetic correlations, \cite{eledopeCPT,KyungSuperAF,Kyung2PSC,KyungCDMFT} 
renormalization of the Fermi surface by interactions, \cite{CivelliCDMFT} or zeros of the Green function. \cite{SakaiImadaPRB} 
\par
The purpose of this paper is to interpret the spectral features of hole-doped and electron-doped systems in a unified manner from the viewpoint of the Mott transition. 
Recently, the spectral features near the Mott transition of the 2D Hubbard model without next-nearest-neighbor hopping have been explained 
by tracing their origins back to the characteristic modes of the one-dimensional (1D) Hubbard model. \cite{Kohno2DHub} 
This paper further extends the argument to the case where next-nearest-neighbor hopping is included, by considering how the spectral weights are shifted by next-nearest-neighbor hopping near the Mott transition, 
and shows that the nature of complicated spectral features in hole-doped and electron-doped systems can be intuitively understood 
by tracing their origins back to the properties of the 1D and 2D Hubbard models without next-nearest-neighbor hopping. 
\par
Based on this picture, various spectral features observed in hole-doped and electron-doped high-$T_c$ cuprates near the Mott transition can be interpreted in a unified manner as properties of the 2D Hubbard model 
whose spectral weights are shifted by next-nearest-neighbor hopping. 
In particular, the different locations of the pseudogap between hole-doped and electron-doped systems \cite{ShenRMP,FlatbandBi2212,LSCO_FS,Ca2NaCuO2Cl2,ArmitageRMP,ArmitagePRL} 
and the disappearance of the spectral weights of the in-gap states toward the Mott transition observed in both hole-doped and electron-doped cuprates \cite{DagottoRMP,ShenRMP,ArmitagePRL,XraySWT_PRB,XraySWT_PRL} 
are naturally explained from the viewpoint of the Mott transition. 
\section{Model and Method}
\label{sec:method}
\subsection{Model} 
We consider the 2D Hubbard model with next-nearest-neighbor hopping defined by the following Hamiltonian. 
\begin{eqnarray}
{\cal H}&=&-t\sum_{\langle i,j\rangle,\sigma}\left(c^\dagger_{i\sigma}c_{j\sigma}+\mbox{H.c.}\right)+t^{\prime}\sum_{\langle\langle i,j\rangle\rangle,\sigma}\left(c^\dagger_{i\sigma}c_{j\sigma}+\mbox{H.c.}\right)\nonumber\\
&&+U\sum_in_{i\uparrow}n_{i\downarrow}-\mu\sum_{i,\sigma}n_{i\sigma},
\label{eq:Hub}
\end{eqnarray}
where $c_{i\sigma}$ and $n_{i\sigma}$ denote the annihilation and number operators, respectively, of an electron at site $i$ with spin $\sigma$. 
The notations $\langle i,j\rangle$ and $\langle\langle i,j\rangle\rangle$ indicate that sites $i$ and $j$ are nearest neighbors and next-nearest neighbors, respectively, on a square lattice. 
We consider the case for $t>0$, $t^{\prime}/t\approx 0.3$, and $U/t\gtrsim 8$, unless otherwise mentioned. 
The hole-doping concentration $\delta$ is defined as $\delta=1-n$, where $n$ denotes the density of electrons. At half-filling ($\delta=0$), the system becomes a Mott insulator 
in the parameter regime we consider in this paper. 
In this paper, the 1D and 2D Hubbard models indicate the models defined by Eq. (\ref{eq:Hub}) without the next-nearest-neighbor hopping term ($t^{\prime}=0$) on a chain and on a square lattice, respectively, unless otherwise mentioned. 
\par
We study the single-particle spectral function $A({\bm k},\omega)$ at zero temperature defined as follows. 
\begin{equation}
A({\bm k},\omega)=\left\{
\begin{array}{rl}
\sum_l\left|\langle l|c^\dagger_{{\bm k}\uparrow}|\mbox{GS}\rangle\right|^2\delta\left(\omega-\varepsilon_l\right)&\quad\mbox{for}\quad\omega>0,\\
\sum_l\left|\langle l|c_{{\bm k}\downarrow}|\mbox{GS}\rangle\right|^2\delta\left(\omega+\varepsilon_l\right)&\quad\mbox{for}\quad\omega<0, 
\end{array}\right.
\label{eq:Akw}
\end{equation}
where $c_{{\bm k}\sigma}$ denotes the annihilation operator of an electron with momentum ${\bm k}$ and spin $\sigma$. 
Here, $\varepsilon_l$ denotes the excitation energy of the eigenstate $|l\rangle$ from the ground state $|\mbox{GS}\rangle$. 
The spectral function $A({\bm k},\omega)$ can also be expressed as 
\begin{equation}
A({\bm k},\omega)=-\frac{1}{\pi}\mbox{Im}G({\bm k},\omega), 
\label{eq:AkwG}
\end{equation}
where $G({\bm k},\omega)$ denotes the retarded single-particle Green function. \cite{ImadaRMP} 
The spectral function $A({\bm k},\omega)$ can be probed by using angle-resolved photoemission spectroscopy. \cite{ShenRMP} 
We consider the properties for $0\le k_y\le k_x\le\pi$ without loss of generality. 
For the 1D and 2D Hubbard models ($t^{\prime}=0$), 
the spectral function in the electron-doped case at $\delta=-\xi$ for $0<\xi\le 1$ is the same as that in the hole-doped case at $\delta=\xi$ 
with $\omega$ and ${\bm k}$ replaced by $-\omega$ and ${\bm \pi}-{\bm k}$, respectively, where ${\bm \pi}$ denotes $\pi$ for 1D and $(\pi,\pi)$ for 2D, 
due to the symmetry under the particle-hole and gauge transformations on bipartite lattices. \cite{TakahashiBook} 
\subsection{Method} 
\label{sec:CPT}
In this paper, cluster perturbation theory (CPT) \cite{CPTPRB,CPTPRL} is employed to calculate $A({\bm k},\omega)$. 
In CPT, the Green function is obtained by connecting cluster Green functions calculated by exact diagonalization through the first-order single-particle hopping process 
without assuming long-range order. 
In the non-interacting case ($U=0$), CPT becomes exact. \cite{CPTPRB,CPTPRL} 
We use cluster Green functions in ($4\times4$)-site clusters, which respect rotation and reflection symmetries of the square lattice. 
\par
Before discussing the results for the 2D Hubbard model with next-nearest-neighbor hopping, 
we confirm the validity of CPT near the Mott transition by comparing the results obtained by using CPT 
with those obtained by using unbiased methods. \cite{Kohno1DHub,Assaad,Furukawa,Poilblanc1h,LiebWu,PreussPG} 
From the viewpoint of numerical calculations, CPT is expected to work better in low-dimensional systems, because boundary effects of clusters are smaller than in high-dimensional systems. 
On the other hand, from the physical viewpoint, CPT is expected to work better in high-dimensional systems, because the free-electron-like mode will be more dominant than in low-dimensional systems. 
In the following, we confirm the validity of CPT for the 1D and 2D Hubbard models near the Mott transition. 
\par
\begin{figure}
\includegraphics[width=8.6cm]{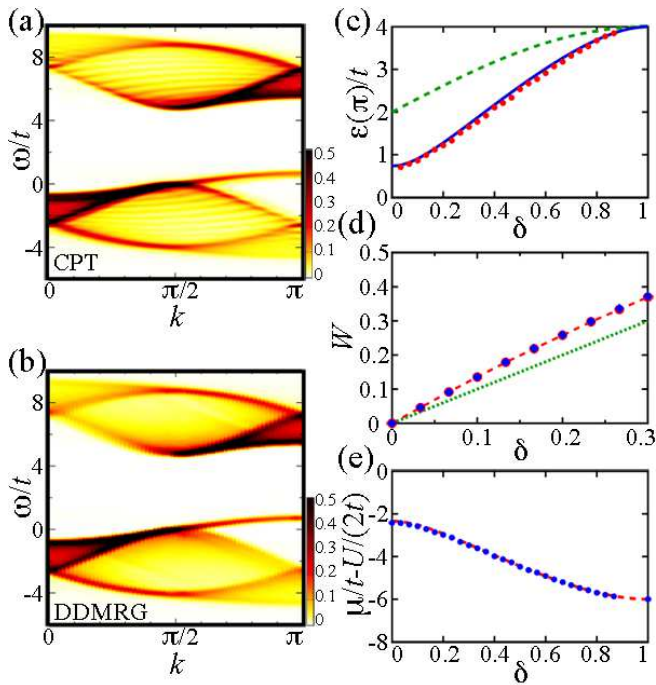}
\caption{Comparisons of CPT results with DDMRG and exact results for the 1D Hubbard model for $U/t=8$. 
(a) $A(k, \omega)t$ obtained by using CPT at $\delta\approx 0.03$. 
(b) The same as (a) but obtained by using the DDMRG method, taken from Ref. \onlinecite{Kohno1DHub}. 
(c) Energy of the mode for $\omega>0$ in the LHB at $k=\pi$ [$\varepsilon(\pi)$] obtained by using CPT (red circles) and 
that obtained by using the Bethe ansatz (solid blue curve). \cite{Kohno1DHub,Kohno2DHub} 
The dashed green curve indicates the energy of the mode at $k=\pi$ for $U = 0$. 
(d) Spectral weight for $\omega > 0$ in the LHB [$W$] 
obtained by using CPT (blue diamonds) and 
that obtained by using the DDMRG method (red circles with dashed line) taken from Ref. \onlinecite{Kohno1DHub}. 
The dotted green line indicates the electron-addition spectral weight in the LHB for $t = 0$. \cite{Eskes} 
(e) Chemical potential $\mu$ obtained by using CPT (blue circles) and that obtained by using the Bethe ansatz (dashed red curve). \cite{LiebWu} 
The values in the $\delta \rightarrow 0$ limit indicate those of $-\Delta/(2t)$ obtained from the Mott gap $\Delta$ between the UHB and the LHB at $\delta = 0$. 
Gaussian broadening is used with standard deviation $\sigma = 0.1t$ for the results obtained by using CPT and the DDMRG method.}
\label{fig:CPT1D}
\end{figure}

First, we compare the results obtained by using 16-site-cluster CPT for the 1D Hubbard model. 
The result for $A(k, \omega)$ of the 1D Hubbard model for $U/t = 8$ near the Mott transition obtained by using CPT [Fig. \ref{fig:CPT1D}(a)] 
agrees reasonably well with that obtained by using the dynamical density-matrix renormalization group (DDMRG) method \cite{DDMRG} [Fig. \ref{fig:CPT1D}(b)]. \cite{Kohno1DHub} 
In particular, the almost symmetric spectral-weight distribution between the upper Hubbard band (UHB) and the lower Hubbard band (LHB) near the Mott transition 
and the emergence of a dispersing mode for $\omega\gtrsim 0$ by hole doping \cite{Kohno1DHub} as well as the characteristic modes, 
such as the spinon mode, the holon mode, and the shadow band, \cite{Kohno1DHub,DDMRGAkw,Shadowband} are well reproduced by CPT. 
In addition, the doping-dependences of the energy of the mode for $\omega>0$ in the LHB at $k=\pi$ [$\varepsilon(\pi)$], 
the spectral weight for $\omega > 0$ in the LHB [$W\left(\equiv\int_{\omega>0\mbox{ in the LHB}} d\omega \int_{-\pi}^{\pi} dkA(k,\omega)/(2\pi)\right)$], 
and the chemical potential $\mu$ [\onlinecite{CPTPRB}] 
obtained by using CPT agree well with the DDMRG results \cite{Kohno1DHub} and the exact results \cite{Kohno1DHub,LiebWu} [Figs. \ref{fig:CPT1D}(c)--\ref{fig:CPT1D}(e)]. 
These comparisons imply that CPT can capture the features near the Mott transition in the 1D Hubbard model for the energy scale of Fig. \ref{fig:CPT1D}. 
\par
\begin{figure}
\includegraphics[width=8.6cm]{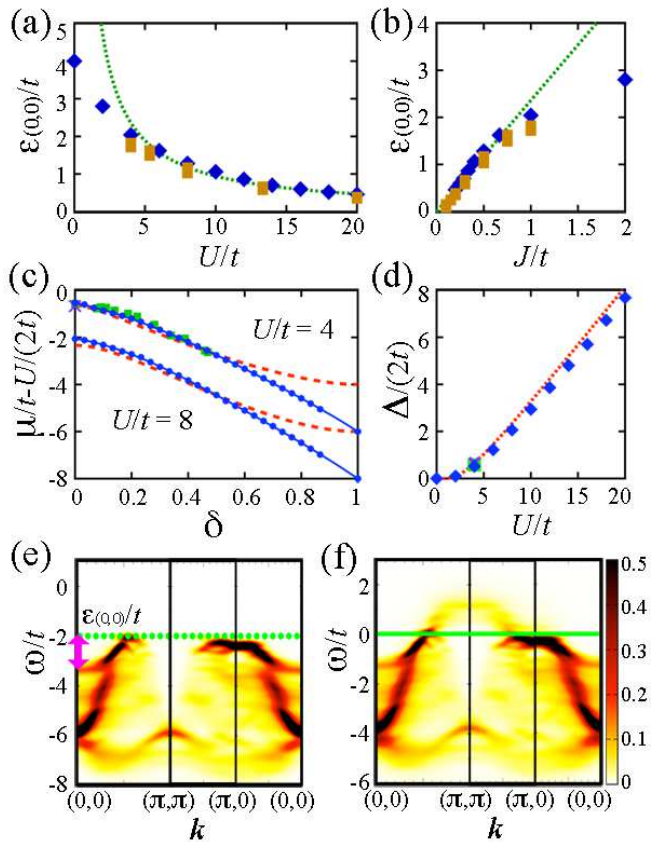}
\caption{Comparisons of CPT results for the 2D Hubbard model ($t^{\prime}=0$) with QMC and exact diagonalization results. 
(a) Bandwidth of the low-$|\omega|$ mode in the LHB in the $(0,0)$--$(\pi,\pi)$ direction at $\delta=0$ [$\varepsilon_{(0,0)}$] [Fig. \ref{fig:CPT2D}(e), pink arrow] (blue diamonds) in the 2D Hubbard model 
obtained by using CPT, taken from Ref. \onlinecite{Kohno2DHub}. 
The dotted green line represents $\sqrt{2}v_{2D}/t$ with the spin-wave velocity of the 2D Heisenberg model $v_{2D}(\approx 1.18\sqrt{2}J$ [\onlinecite{Singh}]) ($J = 4t^2/U$). 
The yellow squares show the bandwidth of the mode of the single-hole excitation in the 2D $t$-$J$ model at $\delta=0$ obtained by using exact diagonalization 
in clusters of up to 26 sites, taken from Ref. \onlinecite{Poilblanc1h}. 
(b) The same as (a) but showing the $J/t$-dependence. 
(c) Chemical potential $\mu$ obtained by using CPT (blue circles with solid lines) and QMC 
(green squares taken from Ref. \onlinecite{Furukawa} and purple cross taken from Ref. \onlinecite{Assaad}) in the 2D Hubbard model. 
The dashed red curves indicate the exact results for the 1D Hubbard model. \cite{LiebWu} 
The upper and lower curves show the results for $U/t = 4$ and 8, respectively. 
The values in the $\delta \rightarrow 0$ limit indicate those of $-\Delta/(2t)$ obtained from the Mott gap $\Delta$ between the UHB and the LHB at $\delta = 0$. 
(d) Mott gap $\Delta$ at $\delta = 0$ obtained by using CPT (blue diamonds taken from Ref. \onlinecite{Kohno2DHub}) and 
QMC (green square taken from Ref. \onlinecite{Furukawa} and purple cross taken from Ref. \onlinecite{Assaad}) in the 2D Hubbard model. 
The dotted red curve indicates the exact results for the 1D Hubbard model. \cite{LiebWu} 
(e) $A({\bm k}, \omega)t$ in the LHB of the 2D Hubbard model for $U/t=8$ at $\delta=0$ obtained by using CPT, taken from Ref. \onlinecite{Kohno2DHub}. 
The dotted green line indicates the $\omega$ value at the top of the LHB. 
The pink arrow indicates the bandwidth of the low-$|\omega|$ mode in the LHB in the $(0,0)$--$(\pi,\pi)$ direction. 
(f) The same as (e) but at $\delta=0.03$, taken from Ref. \onlinecite{Kohno2DHub}. 
The straight solid green line represents $\omega=0$. 
Gaussian broadening is used with standard deviation $\sigma = 0.1t$ for the results obtained by using CPT.}
\label{fig:CPT2D}
\end{figure}

We next compare the results obtained by using ($4\times4$)-site-cluster CPT for the 2D Hubbard model ($t^{\prime}=0$). 
As shown in Figs. \ref{fig:CPT2D}(a) and \ref{fig:CPT2D}(b), the bandwidth of the low-$|\omega|$ mode in the LHB in the $(0,0)$--$(\pi,\pi)$ direction at $\delta=0$ [$\varepsilon_{(0,0)}$] 
[Fig. \ref{fig:CPT2D}(e), pink arrow] obtained by using CPT, \cite{Kohno2DHub} 
which behaves in the large-$U/t$ regime as $\sqrt{2}v_{2D}$, where $v_{2D}$ denotes the spin-wave velocity of the 2D Heisenberg model ($v_{2D}\approx 1.18\sqrt{2}J$ [\onlinecite{Singh}]) ($J = 4t^2/U$), \cite{Kohno2DHub} 
agrees reasonably well \cite{KohnoSpin} with that of the single-hole excitation in the 2D $t$-$J$ model at $\delta=0$ obtained by using exact diagonalization \cite{Poilblanc1h} in the large-$U/t$ regime. 
In addition, the doping-dependence of the chemical potential $\mu$ and the $U/t$-dependence of the Mott gap $\Delta$ at $\delta = 0$ obtained by using CPT 
agree reasonably well with those obtained by quantum Monte Carlo (QMC) simulations for $U/t=4$ [\onlinecite{Furukawa,Assaad}] [Figs. \ref{fig:CPT2D}(c) and \ref{fig:CPT2D}(d)]. 
Furthermore, the overall behavior of $A({\bm k},\omega)$ in the 2D Hubbard model for $U/t=8$ near the Mott transition obtained by using CPT [Fig. \ref{fig:CPT2D}(f)] \cite{Kohno2DHub} 
agrees reasonably well with that obtained by a QMC simulation in Ref. \onlinecite{PreussPG}. 
These comparisons imply that CPT can capture the features near the Mott transition in the 2D Hubbard model for the energy scale of Fig. \ref{fig:CPT2D}. 
\par
The doping-dependence of the chemical potential near the Mott transition in the 2D Hubbard model appears similar to that in the 1D Hubbard model [Fig. \ref{fig:CPT2D}(c)]. 
To judge whether the value of the critical exponent of the charge susceptibility $\chi_c(=\partial n/\partial \mu)$ for the Mott transition in 2D \cite{ImadaRMP,Furukawa,Assaad,KohnotJ} is 
the same as that in 1D, \cite{1DHubChic} more careful analyses than those in the present study, such as size-scaling analyses, might be necessary. 
In any case, the characteristic features of the Mott transition shown in Ref. \onlinecite{Kohno2DHub} and in this paper do not depend on the value of the exponent. 
\par
Because the results shown in this paper as well as in Ref. \onlinecite{Kohno2DHub} are based on ($4\times4$)-site cluster Green functions, 
the properties in the very small-$|\omega|$ regime as well as the details of lineshapes and fine structures of the spectral function might not be well resolved due to finite-size effects. 
This paper as well as Ref. \onlinecite{Kohno2DHub} discusses overall spectral features for which the ($4\times4$)-site-cluster CPT appears to give reliable results. 
\par
Here, some technical details for how CPT was implemented in this paper and in Ref. \onlinecite{Kohno2DHub} are mentioned. 
The Lanczos algorithm was used for the calculation of cluster Green functions in clusters without ground-state magnetization. 
The Green functions were calculated by considering superclusters so the electron density $n$ could be controlled arbitrarily. \cite{CPTPRB} 
The CPT results with Lorentzian broadening (half-width at half-maximum $\eta=0.16t$) were deconvolved into those with Gaussian broadening (standard deviation $\sigma=0.1t$). 
The chemical potential $\mu$ was set such that the spectral weight for $\omega < 0$ was equal to $n/2$: $\int_{-\infty}^0d\omega\int d{\bm k}A({\bm k}, \omega)/(2\pi)^d=n/2$ in $d$ dimensions. \cite{CPTPRB} 
At half-filling, it was set at $U/2$. 
Since $A({\bm k}, \omega)$ was calculated at discrete values of $\omega$, spaced by intervals of $\Delta\omega = 0.02t$, 
the position of $\omega=0$ was generally located between the discrete calculation points. 
In this paper and Ref. \onlinecite{Kohno2DHub}, $A({\bm k}, \omega)$ at the discrete $\omega$-point closest to $\omega = 0$ for $\omega < 0$ 
is shown as $A({\bm k}, \omega)$ for $\omega\approx 0$. 
It was confirmed that the $\omega$ discretization was sufficiently fine such that there was no serious difference within the interval around $\omega = 0$ 
for the discussions in this paper and Ref. \onlinecite{Kohno2DHub}. 
\par
In this paper as well as in Refs. \onlinecite{Kohno2DHub} and \onlinecite{Kohno1DHub}, 
peaks [and edges of characteristic continua (apart from Gaussian broadening)] of the spectral function are referred to as modes. 
The $\omega$ value at the top of the LHB at $\delta = 0$ in the 2D Hubbard model is defined 
as the highest value of $\omega$ of the high-$\omega$ mode in the $(0,0)$--$(\pi,\pi)$ direction in the LHB [Fig. \ref{fig:CPT2D}(e), dotted green line]. 
\section{Spectral features of the 2D Hubbard model with next-nearest-neighbor hopping}
\label{sec:features}
\subsection{CPT results}
\label{sec:2ndNN}
\begin{figure*}
\includegraphics[width=17.5cm]{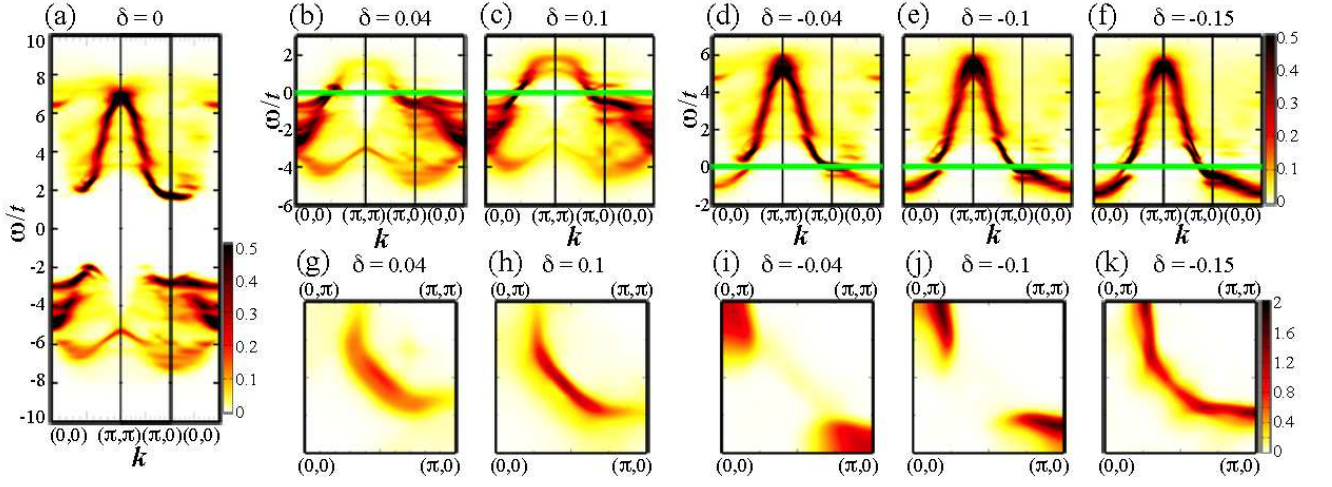}
\caption{(a)--(f) $A({\bm k},\omega)t$ of the 2D Hubbard model with next-nearest-neighbor hopping for $U/t=8$ and $t^{\prime}/t=0.3$ 
at $\delta=0$, $0.04$, $0.1$, $-0.04$, $-0.1$, and $-0.15$ in the LHB [(a)-(c)] and the UHB [(a), (d)--(f)], obtained by using CPT. 
The solid green lines represent $\omega=0$. 
(g)--(k) The same as in (b)--(f) but for $\omega\approx 0$. Gaussian broadening is used with standard deviation $\sigma = 0.1t$.}
\label{fig:2ndNN}
\end{figure*}
The results obtained by using CPT for the 2D Hubbard model with next-nearest-neighbor hopping for $U/t=8$ and $t^{\prime}/t=0.3$ are shown in Fig. \ref{fig:2ndNN}. 
At half-filling, the bottom of the UHB is located near $(\pi,0)$, whereas the top of the LHB is located near $(\pi/2,\pi/2)$ [Fig. \ref{fig:2ndNN}(a)]. 
In hole-doped systems near the Mott transition, the spectral weights for $\omega\approx 0$ are primarily located around $(\pi/2,\pi/2)$, 
and pseudogap behavior appears around $(\pi,0)$ and $(0,\pi)$ [Figs. \ref{fig:2ndNN}(b), \ref{fig:2ndNN}(c), \ref{fig:2ndNN}(g), and \ref{fig:2ndNN}(h)]. 
These features are essentially the same as those of the 2D Hubbard model [Fig. \ref{fig:CPT2D}(f)] \cite{Kohno2DHub} 
except that the pseudogap behavior is enhanced by next-nearest-neighbor hopping. 
In contrast, in electron-doped systems near the Mott transition, the spectral weights for $\omega\approx 0$ are primarily located around $(\pi,0)$ and $(0,\pi)$, 
and pseudogap behavior appears around $(\pi/2,\pi/2)$ [Figs. \ref{fig:2ndNN}(d), \ref{fig:2ndNN}(e), \ref{fig:2ndNN}(i), and \ref{fig:2ndNN}(j)]. 
The spectral weights of the mode for $\omega<0$ around $(0,0)$ in the UHB disappear toward the Mott transition in electron-doped systems [Figs. \ref{fig:2ndNN}(a) and \ref{fig:2ndNN}(d)--\ref{fig:2ndNN}(f)] 
as do those for $\omega>0$ around $(\pi,\pi)$ in the LHB in hole-doped systems [Figs. \ref{fig:CPT2D}(e), \ref{fig:CPT2D}(f), and \ref{fig:2ndNN}(a)--\ref{fig:2ndNN}(c)]. \cite{Kohno2DHub,Kohno1DHub} 
The dispersion relation around $(\pi,0)$ is considerably flat near the Mott transition in both hole-doped and electron-doped systems [Figs. \ref{fig:CPT2D}(e), \ref{fig:CPT2D}(f), and \ref{fig:2ndNN}(a)--\ref{fig:2ndNN}(f)]. \cite{Kohno2DHub} 
By increasing the doping concentration of electrons (holes), the pseudogap decreases in electron-doped (hole-doped) systems and closes at some value of the doping concentration, 
which results in a Fermi surface similar to that of the non-interacting case [Figs. \ref{fig:2ndNN}(b)--\ref{fig:2ndNN}(k)]. \cite{Kohno2DHub} 
Spectral features similar to the above have been observed in high-$T_c$ cuprates. \cite{ShenRMP,ArmitageRMP,ArmitagePRL,Ca2NaCuO2Cl2,LSCO_FS,FlatbandBi2212,UniversalFlatbandBi2212,DagottoRMP} 
\par
Although similar spectral features have been obtained in the 2D Hubbard and $t$-$J$ models with next-nearest-neighbor hopping by various methods, their interpretations are controversial. 
\cite{KusunoseWeakU,ModeCouping,KyungSuperAF,Kyung2PSC,KyungCDMFT,eledopeCPT,eledopeVCPT,CivelliCDMFT,AichhornVCA,eledopeDCA,tJTohyamaMaekawa,tJTohyama,tJPrelovsek,eledopeFLEX,eledopeSBtJ,eledopeSBHub,SakaiImadaPRB,DahnkenVCPT,FS_DCA} 
For the pseudogap around $(\pi/2,\pi/2)$ in electron-doped systems, there are several interpretations, such as that it is the result of the band splitting caused by antiferromagnetic long-range order, \cite{KuskoMF,KusunoseWeakU,ModeCouping,AichhornVCA,eledopeDCA,eledopeFLEX,eledopeSBtJ,eledopeSBHub,tJTohyamaMaekawa,tJTohyama,tJPrelovsek} short-range antiferromagnetic correlations, \cite{eledopeCPT,KyungSuperAF,Kyung2PSC,KyungCDMFT} 
renormalization of the Fermi surface by interactions, \cite{CivelliCDMFT} or zeros of the Green function. \cite{SakaiImadaPRB} 
\subsection{Spectral-weight shift caused by next-nearest-neighbor hopping} 
\label{sec:RPAtd}
\begin{figure}
\includegraphics[width=8.5cm]{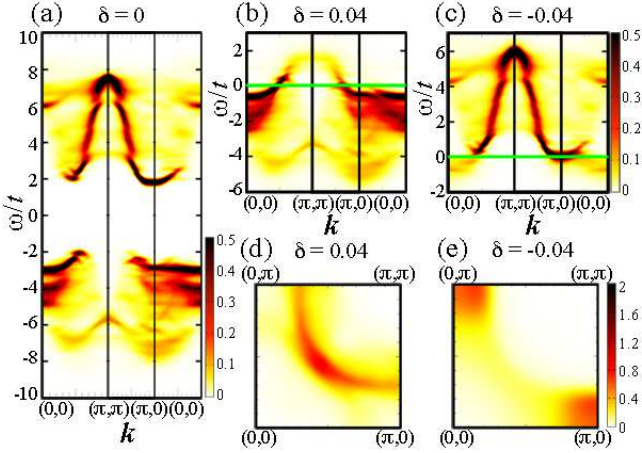}
\caption{$A({\bm k},\omega)t$ obtained by using the RPA-type approximation for next-nearest-neighbor hopping [Eq. (\ref{eq:RPAtd})] for $t^{\prime}/t=0.5$. 
The CPT results for the 2D Hubbard model ($U/t=8$ and $t^{\prime}=0$) \cite{Kohno2DHub} are used as $G_{2D}({\bm k},\omega)$ in Eq. (\ref{eq:RPAtd}). 
(a)--(c) $A({\bm k},\omega)t$ at $\delta=0$, $0.04$, and $-0.04$ in the LHB [(a), (b)] and the UHB[(a), (c)]. 
The solid green lines represent $\omega=0$. 
(d) and (e) The same as in (b) and (c) but for $\omega\approx 0$. Gaussian broadening is used with standard deviation $\sigma = 0.1t$.}
\label{fig:RPAtd}
\end{figure}
Here, we interpret the spectral features as a result of the spectral-weight shift caused by next-nearest-neighbor hopping near the Mott transition. 
In the non-interacting case ($U=0$), since the dispersion relation of the single-particle excitation is obtained as $\varepsilon({\bm k})=-2t(\cos k_x+\cos k_y)+4t^{\prime}\cos k_x\cos k_y-\mu$, 
the next-nearest-neighbor hopping ($t^{\prime}$ term) shifts the spectral weights to higher and lower values of $\omega$ in the momentum regime 
where the values of the Fourier transform of the next-nearest-neighbor hopping integral, $t^{\prime}({\bm k})[\equiv 4t^{\prime}\cos k_x\cos k_y]$, are positive and negative, respectively. 
Similarly, in the interacting case ($U>0$), the next-nearest-neighbor hopping is expected to shift the spectral weights of the 2D Hubbard model 
to higher and lower values of $\omega$ for $t^{\prime}({\bm k})>0$ and $t^{\prime}({\bm k})<0$, respectively. \cite{SakaiImadaPRB,SakaiImada} 
The shift should also depend on the spectral-weight distribution. 
A simple approximation that exhibits such spectral-weight shift is the following. 
In the small-$|t^{\prime}/t|$ regime, by considering next-nearest-neighbor hopping as perturbation, 
the Green function of the 2D Hubbard model with next-nearest-neighbor hopping can be approximated \cite{NoteForPerturbation} as 
\begin{equation}
G({\bm k},\omega)\approx\frac{G_{2D}({\bm k},\omega)}{1-t^{\prime}({\bm k})G_{2D}({\bm k},\omega)}, 
\label{eq:RPAtd}
\end{equation}
analogously to the random-phase approximation (RPA) for interchain hopping from the 1D limit \cite{RPAWen,RPASchulz,RPAEsslerSpin,RPAEssler,ArrigoniRPA,RPABoies,RPAClarke,RPATsvelik,RPASchenhammer,KohnoNP,KohnoQ1DHeisH,RPARibeiro,Kohno2DHub} 
and that for intersite hopping from the atomic limit. \cite{HubbardRPA,MetznerRPA} 
Here, $G_{2D}({\bm k},\omega)$ denotes the Green function of the 2D Hubbard model ($t^{\prime}=0$). 
Then, in the momentum regime where the values of $t^{\prime}({\bm k})$ are positive and negative, 
the spectral weights are shifted to higher and lower values of $\omega$, respectively, by the perturbation (next-nearest-neighbor hopping). \cite{KohnoNP,KohnoQ1DHeisH} 
The shift becomes large, where the value of $|t^{\prime}({\bm k})|$ and the spectral weight are large. \cite{KohnoNP} 
This approximation becomes exact for $U=0$ and recovers the properties of the 2D Hubbard model [$G_{2D}({\bm k},\omega)$] for $t^{\prime}=0$. 
\par
Figure \ref{fig:RPAtd} shows the results obtained by using Eq. (\ref{eq:RPAtd}) for $t^{\prime}/t=0.5$, 
where the CPT results for the 2D Hubbard model ($U/t=8$ and $t^{\prime}=0$) \cite{Kohno2DHub} are used as $G_{2D}({\bm k},\omega)$. 
At half-filling, the bottom of the UHB is located near $(\pi,0)$, whereas the top of the LHB is located near $(\pi/2,\pi/2)$ [Fig. \ref{fig:RPAtd}(a)], as in the case of the CPT results [Fig. \ref{fig:2ndNN}(a)]. This feature can be explained as follows. 
Because the values of $t^{\prime}({\bm k})$ are negative and large in magnitude around $(\pi,0)$ and $(0,\pi)$ [$t^{\prime}({\bm k})\approx -4t^{\prime}<0$], 
the flat mode carrying large spectral weights around $(\pi,0)$ and $(0,\pi)$ is shifted considerably to lower values of $\omega$ by next-nearest-neighbor hopping, 
whereas the spectral weights around $(\pi/2,\pi/2)$ remain almost unaffected by next-nearest-neighbor hopping [Eq. (\ref{eq:RPAtd}), $t^{\prime}({\bm k})\approx 0$]. 
Thus, in the UHB, the flat mode around $(\pi,0)$, which was originally located above the bottom of the UHB near $(\pi/2,\pi/2)$ in the 2D Hubbard model [Fig. \ref{fig:CPT2D}(e) in the case of the UHB], 
is shifted down below the states around $(\pi/2,\pi/2)$, 
which causes large spectral weights for $\omega\approx 0$ around $(\pi,0)$ and $(0,\pi)$ rather than around $(\pi/2,\pi/2)$ in an electron-doped system near the Mott transition [Figs. \ref{fig:RPAtd}(c) and \ref{fig:RPAtd}(e)]. 
In the LHB, the flat mode around $(\pi,0)$, which was originally located below the top of the LHB near $(\pi/2,\pi/2)$ in the 2D Hubbard model [Fig. \ref{fig:CPT2D}(e)], is shifted to further lower values of $\omega$, 
with the states around $(\pi/2,\pi/2)$ almost unchanged. 
Thus, the spectral weights for $\omega\approx 0$ near the Mott transition in a hole-doped system remain primarily located around $(\pi/2,\pi/2)$ 
with the pseudogap behavior around $(\pi,0)$ and $(0,\pi)$ enhanced [Figs. \ref{fig:RPAtd}(b) and \ref{fig:RPAtd}(d)]. 
\par
As for the energy scale of the pseudogap in an electron-doped system near the Mott transition, 
because the flat mode around $(\pi,0)$ is shifted to lower values of $\omega$ by next-nearest-neighbor hopping with the states around $(\pi/2,\pi/2)$ almost unaffected [Eq. (\ref{eq:RPAtd})], 
the energy difference between the lower edge of the continuum bending back near $(\pi/2,\pi/2)$ and the flat mode around $(\pi,0)$ in the UHB 
is primarily determined by the shift of the flat mode of the order of $t^{\prime}$, $O(t^{\prime})$, in the large-$U/t$ regime [Figs. \ref{fig:2ndNN}(a), \ref{fig:2ndNN}(d), \ref{fig:2ndNN}(e), \ref{fig:RPAtd}(a), and \ref{fig:RPAtd}(c)]. 
In a hole-doped system near the Mott transition, due to the same reason, the energy difference between the upper edge of the continuum bending back near $(\pi/2,\pi/2)$ and the flat mode around $(\pi,0)$ in the LHB is enhanced by next-nearest-neighbor hopping. 
Thus, the energy scale of the pseudogap defined by this energy difference, which almost coincides with the pseudogap defined as the energy difference between the Fermi level ($\omega=0$) and the main peak of $\rho(\omega)[\equiv\int d{\bm k}A({\bm k},\omega)/(2\pi)^2$] in the $\delta\rightarrow +0$ limit, \cite{Kohno2DHub} 
is primarily determined by the shift of the flat mode of $O(t^{\prime})$ 
in addition to the effect of the antiferromagnetic fluctuation of the order of $J(=4t^2/U)$ in the large-$U/t$ regime \cite{Kohno2DHub} [Figs. \ref{fig:2ndNN}(a)--\ref{fig:2ndNN}(c), \ref{fig:RPAtd}(a), and \ref{fig:RPAtd}(b)]. 
\par
The argument on the spectral-weight shift caused by next-nearest-neighbor hopping can be straightforwardly generalized to the cases where there are further-neighbor hoppings, 
by defining $t^{\prime}({\bm k})$ in Eq. (\ref{eq:RPAtd}) such that the Fourier transforms of the further-neighbor hopping integrals are included. 
In the momentum regime where the values of $t^{\prime}({\bm k})$ are positive and negative, the spectral weights are shifted to higher and lower values of $\omega$, respectively, 
by the perturbation (next-nearest- and further-neighbor hoppings). 
\par
The above RPA-type argument reasonably well explains the overall spectral features, such as the locations of large spectral weights and the pseudogap behavior (reduction in spectral weight) 
for $\omega\approx 0$ near the Mott transition, for the 2D Hubbard model with next-nearest-neighbor hopping obtained by using CPT [Figs. \ref{fig:2ndNN}(b)--\ref{fig:2ndNN}(e) and \ref{fig:2ndNN}(g)--\ref{fig:2ndNN}(j)] 
and those observed in high-$T_c$ cuprates. \cite{ArmitageRMP,ArmitagePRL,FlatbandBi2212,Ca2NaCuO2Cl2,ShenRMP,LSCO_FS,UniversalFlatbandBi2212,DagottoRMP} 
However, the spectral weights for $\omega\approx 0$ in the pseudogap momentum regime near the Mott transition obtained by using CPT [Figs. \ref{fig:2ndNN}(g)--\ref{fig:2ndNN}(j)] 
are considerably smaller than those of the RPA-type approximation [Figs. \ref{fig:RPAtd}(d) and \ref{fig:RPAtd}(e)], 
which implies that processes omitted in the RPA-type approximation [Eq. (\ref{eq:RPAtd})] enhance the pseudogap behavior. 
The spectral weights for $\omega\approx 0$ in the pseudogap momentum regime observed in high-$T_c$ cuprates are also considerably reduced 
near the Mott transition. \cite{ArmitageRMP,ArmitagePRL,FlatbandBi2212,Ca2NaCuO2Cl2,ShenRMP,LSCO_FS} 
\par
From the viewpoint of the spectral-weight shift, the strong reduction in spectral weight for $\omega\approx 0$ in the pseudogap momentum regime near the Mott transition 
[Figs. \ref{fig:2ndNN}(b)--\ref{fig:2ndNN}(e) and \ref{fig:2ndNN}(g)--\ref{fig:2ndNN}(j)] \cite{ArmitageRMP,ArmitagePRL,FlatbandBi2212,Ca2NaCuO2Cl2,ShenRMP,LSCO_FS} 
can be interpreted as a separation of the modes due to the large shift of the flat mode around $(\pi,0)$ caused by next-nearest-neighbor hopping. 
In electron-doped systems near the Mott transition, the mode fading away toward the Mott transition can be effectively separated around $\omega=0$ 
from the continuum for $\omega>0$ around $(\pi/2,\pi/2)$ as a result of the large downward shift of the flat mode around $(\pi,0)$ [Figs. \ref{fig:2ndNN}(d) and \ref{fig:2ndNN}(e)]. 
In hole-doped systems near the Mott transition, the flat mode can be almost separated around $\omega=0$ 
from the mode fading away toward the Mott transition due to the downward shift of the flat mode around $(\pi,0)$ [Figs. \ref{fig:2ndNN}(b) and \ref{fig:2ndNN}(c)]. 
The strong reduction in spectral weight for $\omega\approx 0$ in the pseudogap momentum regime near the Mott transition [Figs. \ref{fig:2ndNN}(b)--\ref{fig:2ndNN}(e) and \ref{fig:2ndNN}(g)--\ref{fig:2ndNN}(j)] 
might also be related to antiferromagnetic fluctuations, pairing fluctuations, or other effects suggested in the literature. 
\cite{AndersonRVB,DagottoRMP,NagaosaRMP,YanasePG,ImadaRMP,SakaiImadaPRB,KuskoMF,KusunoseWeakU,ModeCouping,KyungSuperAF,Kyung2PSC,KyungCDMFT,eledopeCPT,eledopeVCPT,CivelliCDMFT,AichhornVCA,eledopeDCA,tJTohyamaMaekawa,tJTohyama,tJPrelovsek,eledopeFLEX,eledopeSBtJ,eledopeSBHub} 
For details, further investigations, including careful analyses on finite-size effects in CPT, are necessary. 
\subsection{Interpretation of spectral features of electron-doped high-$T_c$ cuprates} 
Based on the above results for the 2D Hubbard model with next-nearest-neighbor hopping, the spectral features observed in electron-doped high-$T_c$ cuprates \cite{ArmitageRMP,ArmitagePRL,ShenRMP} 
can be naturally explained from the viewpoint of the Mott transition 
as properties of the 2D Hubbard model whose spectral weights are shifted by next-nearest-neighbor hopping. 
The states referred to as in-gap states or doping-induced states can be identified as the mode fading away toward the Mott transition \cite{Kohno2DHub} whose spectral weights are shifted by next-nearest-neighbor hopping. 
The pseudogap around $(\pi/2,\pi/2)$ can be primarily interpreted as a reduction in spectral weight for $\omega\approx 0$ around $(\pi/2,\pi/2)$ due to the spectral-weight shift of the flat mode around $(\pi,0)$ caused by next-nearest-neighbor hopping: 
the next-nearest-neighbor hopping shifts the flat mode carrying large spectral weights around $(\pi,0)$ down below $\omega=0$, which lowers the chemical potential, 
leading to a reduction in spectral weight for $\omega\approx 0$ around $(\pi/2,\pi/2)$ with the doping-induced states existing below $\omega=0$. 
\par
In this picture, the antiferromagnetic long-range order is not necessary for the pseudogap behavior. 
The pseudogap behavior is explained as a result of the spectral features near the Mott transition [the presence of the mode fading away toward the Mott transition and 
the flat mode carrying large spectral weights around $(\pi,0)$] and the spectral-weight shift caused by next-nearest-neighbor hopping. 
This picture explains not only the pseudogap behavior (reduction in spectral weight) around $(\pi/2,\pi/2)$ in electron-doped systems 
[Figs. \ref{fig:2ndNN}(i), \ref{fig:2ndNN}(j), and \ref{fig:RPAtd}(e)] \cite{ArmitageRMP,ArmitagePRL,ShenRMP} 
and that around $(\pi,0)$ and $(0,\pi)$ in hole-doped systems [Figs. \ref{fig:2ndNN}(g), \ref{fig:2ndNN}(h), and \ref{fig:RPAtd}(d)] \cite{ShenRMP,LSCO_FS,FlatbandBi2212,Ca2NaCuO2Cl2} 
but also the doping-induced states \cite{ArmitagePRL,ShenRMP,DagottoRMP,XraySWT_PRB,XraySWT_PRL} 
and the flat dispersion relation around $(\pi,0)$ \cite{ShenRMP,FlatbandBi2212,UniversalFlatbandBi2212} [Figs. \ref{fig:2ndNN}(b)--\ref{fig:2ndNN}(f), \ref{fig:RPAtd}(b), and \ref{fig:RPAtd}(c)] in a unified manner. 
In addition, the energy scale of the Mott gap is of the order of $U$ in the large-$U/t$ regime in this picture. 
\section{Discussion and summary} 
\label{sec:summary}
\begin{figure*}
\includegraphics[width=15cm]{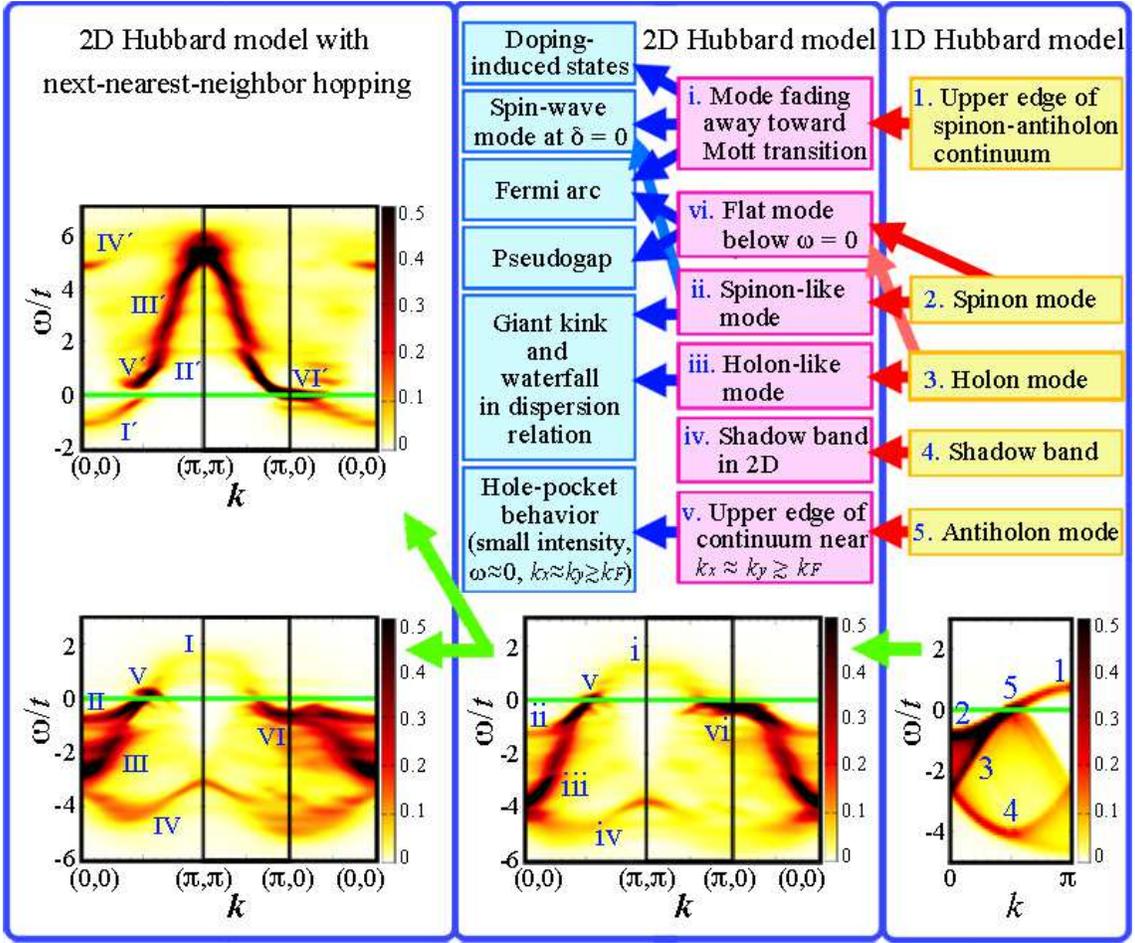}
\caption{Relationship between spectral features and modes near the Mott transition. 
(Left column) $A({\bm k},\omega)t$ of the UHB in the electron-doped case ($\delta=-0.03$) (upper panel) and 
that of the LHB in the hole-doped case ($\delta=0.03$) (lower panel) 
in the 2D Hubbard model with next-nearest-neighbor hopping ($U/t=8$ and $t^{\prime}/t=0.3$) obtained by using CPT. 
The blue numbers I--VI and I$^{\prime}$--VI$^{\prime}$ indicate the modes 
whose origins in the 2D Hubbard model ($t^{\prime}=0$) are the modes indicated by the blue numbers i--vi on the intensity plot in the center column. 
(Center column) The blue arrows indicate which modes are responsible for the spectral features in the 2D Hubbard model. \cite{Kohno2DHub} 
The red arrows indicate which modes of the 1D Hubbard model are the primary origins of the modes of the 2D Hubbard model. \cite{Kohno2DHub} 
Here, $k_F$ denotes the value of $k_x(=k_y)$ of the Fermi surface in the $(0,0)$--$(\pi,\pi)$ direction. 
The intensity plot shows $A({\bm k},\omega)t$ of the LHB at $\delta=0.03$ in the 2D Hubbard model ($U/t=8$ and $t^{\prime}=0$) 
obtained by using CPT [Fig. \ref{fig:CPT2D}(f)] taken from Ref. \onlinecite{Kohno2DHub}. 
The blue numbers i--vi on the intensity plot indicate the modes described in the above diagram. \cite{Kohno2DHub} 
(Right column) The intensity plot shows $A(k,\omega)t$ of the LHB at $\delta\approx 0.03$ in the 1D Hubbard model ($U/t=8$) obtained by using the DDMRG method 
[Fig. \ref{fig:CPT1D}(b)] taken from Ref. \onlinecite{Kohno1DHub}. 
The blue numbers 1--5 on the intensity plot indicate the modes described in the above diagram. \cite{Kohno1DHub} 
The solid green lines on the intensity plots represent $\omega=0$. Gaussian broadening is used with standard deviation $\sigma = 0.1t$.}
\label{fig:modes}
\end{figure*}

Various spectral features observed in hole-doped and electron-doped high-$T_c$ 
cuprates \cite{DagottoRMP,ShenRMP,Graf,UniversalFlatbandBi2212,FlatbandBi2212,LSCO_FS,holepocketYBCO,ArmitageRMP,ArmitagePRL,Ca2NaCuO2Cl2,XraySWT_PRB,XraySWT_PRL} 
can be explained in a unified manner as properties of the 2D Hubbard model with next-nearest-neighbor hopping, whose origins can be traced back to the characteristic modes of the 2D Hubbard model 
based on the consideration of how the next-nearest-neighbor hopping shifts the spectral weights (Fig. \ref{fig:modes}). 
The origins of the modes of the 2D Hubbard model can be further traced back to those of the 1D Hubbard model (Fig. \ref{fig:modes}). \cite{Kohno2DHub} 
\par
Specifically, the spectral features in hole-doped and electron-doped systems with next-nearest-neighbor hopping are characterized 
by the modes indicated by the blue numbers I--VI and I$^{\prime}$--VI$^{\prime}$ on the intensity plots in the left column of Fig. \ref{fig:modes}. 
These modes are obtained as a result of the spectral-weight shift caused by next-nearest-neighbor hopping from the modes of the 2D Hubbard model indicated by the blue numbers i--vi in the center column of Fig. \ref{fig:modes}. 
The origins of the modes of the 2D Hubbard model (Fig. \ref{fig:modes}, modes i--vi) can be traced back to the modes of the 1D Hubbard model 
indicated by the blue number 1--5 in the right column of Fig. \ref{fig:modes}, 
by considering the spectral-weight shift caused by interchain hopping. \cite{Kohno2DHub} 
Based on these relationships, various spectral features can be interpreted as follows. 
\par
{\it Doping-induced states.--}
The mode for $\omega>0$ in the LHB in hole-doped systems (Fig. \ref{fig:modes}, modes i and I) and 
that for $\omega<0$ around $(0,0)$ in the UHB in electron-doped systems (Fig. \ref{fig:modes}, mode I$^{\prime}$ and mode i in the electron-doped case) 
correspond to the states referred to as in-gap states or doping-induced states. \cite{Eskes,DagottoDOS,SakaiImadaPRB,SakaiImada,PhillipsRMP,Liebsch,DagottoRMP,ShenRMP,ArmitagePRL,XraySWT_PRB,XraySWT_PRL} 
These modes originate from the mode of the upper edge of the spinon-antiholon continuum in the 1D Hubbard model (Fig. \ref{fig:modes}, mode 1). \cite{Kohno2DHub} 
All these modes (Fig. \ref{fig:modes}, modes 1, i, I, and I$^{\prime}$) remain dispersing until the Mott transition occurs and fade away toward the Mott transition. \cite{Kohno1DHub,Kohno2DHub} 
\par
{\it Fermi arc and pseudogap in hole-doped systems.--}
In hole-doped systems near the Mott transition, the spectral weights for $\omega\approx 0$ are primarily located around $(\pi/2,\pi/2)$ 
[Fig. \ref{fig:modes}, lower panels in the left and center columns; Figs. \ref{fig:2ndNN}(g) and \ref{fig:2ndNN}(h)]. \cite{Kohno2DHub} 
There are essentially no spectral weights for $\omega\approx 0$ in the $(0,0)$--$(\pi,0)$ direction, because the flat mode around $(\pi,0)$ (Fig. \ref{fig:modes}, modes vi and VI) is located below $\omega=0$. \cite{Kohno2DHub} 
The spectral weights for $\omega\approx 0$ in the $(\pi,0)$--$(\pi,\pi)$ direction reduce significantly near the Mott transition 
due to the property of the mode fading away toward the Mott transition (Fig. \ref{fig:modes}, modes i and I). \cite{Kohno2DHub} 
In addition, the flat mode (Fig. \ref{fig:modes}, mode VI) can be almost separated around $\omega=0$ 
from the mode fading away toward the Mott transition (Fig. \ref{fig:modes}, mode I) 
due to the downward shift of the flat mode around $(\pi,0)$ caused by next-nearest-neighbor hopping [Eq. (\ref{eq:RPAtd}), $t^{\prime}({\bm k})\approx -4t^{\prime}<0$]. 
Because of these reductions in spectral weight for $\omega\approx 0$ in the $(0,0)$--$(\pi,0)$ and $(\pi,0)$--$(\pi,\pi)$ directions, the Fermi arc behavior appears in hole-doped systems 
[Figs. \ref{fig:2ndNN}(g) and \ref{fig:2ndNN}(h)]. \cite{Kohno2DHub} 
The pseudogap defined as the energy difference between the Fermi level ($\omega=0$) and the main peak of $\rho(\omega)$ is primarily determined 
by the energy difference between the Fermi level and the flat mode around $(\pi,0)$ 
(Fig. \ref{fig:modes}, modes vi and VI) in hole-doped systems. \cite{Kohno2DHub} 
The pseudogap is enhanced by next-nearest-neighbor hopping in hole-doped systems, because the flat mode around $(\pi,0)$ is shifted downward by next-nearest-neighbor hopping [Fig. \ref{fig:modes}, mode VI; Eq. (\ref{eq:RPAtd})]. 
\par
{\it Pseudogap around $(\pi/2,\pi/2)$ in electron-doped systems.--}
In electron-doped systems with next-nearest-neighbor hopping near the Mott transition, the spectral weights for $\omega\approx 0$ are primarily located around $(\pi,0)$ 
[Fig. \ref{fig:modes}, upper panel in the left column; Figs. \ref{fig:2ndNN}(i) and \ref{fig:2ndNN}(j)], which are due to the flat mode around $(\pi,0)$ (Fig. \ref{fig:modes}, mode VI$^{\prime}$). 
Because the flat mode around $(\pi,0)$ carrying large spectral weights, which was originally located above $\omega=0$ (Fig. \ref{fig:modes}, mode vi in the electron-doped case) in the UHB of the 2D Hubbard model, 
is significantly shifted by next-nearest-neighbor hopping (Fig. \ref{fig:modes}, mode VI$^{\prime}$) down below the lower edge of the continuum around $(\pi/2,\pi/2)$ (Fig. \ref{fig:modes}, mode V$^{\prime}$), 
the flat mode is located almost at $\omega=0$ in the small-electron-doping regime, which causes large spectral weights for $\omega\approx 0$ around $(\pi,0)$. 
The spectral weights for $\omega\approx 0$ around $(\pi/2,\pi/2)$ reduce significantly near the Mott transition 
due to the property of the mode fading away toward the Mott transition (Fig. \ref{fig:modes}, mode I$^{\prime}$). 
In addition, the mode fading away toward the Mott transition (Fig. \ref{fig:modes}, mode I$^{\prime}$) can be effectively separated around $\omega=0$ 
from the continuum around $(\pi/2,\pi/2)$ whose lower edge can be identified as the mode (Fig. \ref{fig:modes}, mode V$^{\prime}$) originating from the 1D antiholon mode (Fig. \ref{fig:modes}, mode 5) 
as a result of the large downward shift of the flat mode around $(\pi,0)$ caused by next-nearest-neighbor hopping [Fig. \ref{fig:modes}, mode VI$^{\prime}$; Eq. (\ref{eq:RPAtd})]. 
The pseudogap can be explained as the reduction in spectral weight for $\omega\approx 0$ around $(\pi/2,\pi/2)$ with the doping-induced states (Fig. \ref{fig:modes}, mode I$^{\prime}$) existing below $\omega=0$. 
\par
{\it Giant kink and waterfall.--}
The giant kink and the waterfall behavior in the $(0,0)$--$(\pi,\pi)$ direction are explained 
as a result of the mode (Fig. \ref{fig:modes}, modes ii, II, and II$^{\prime}$) originating from the 1D spinon mode (Fig. \ref{fig:modes}, mode 2) 
and the mode (Fig. \ref{fig:modes}, modes iii, III, and III$^{\prime}$) originating from the 1D holon mode (Fig. \ref{fig:modes}, mode 3). \cite{Kohno2DHub} 
\par
{\it Hole-pocket behavior in hole-doped systems.--}
The signature for the hole-pocket behavior in hole-doped systems is explained as a result of the upper edge of the continuum [bending back near $(\pi/2,\pi/2)$] (Fig. \ref{fig:modes}, modes v and V). \cite{Kohno2DHub} 
The upper edge around $(\pi/2,\pi/2)$ originates from the 1D antiholon mode (Fig. \ref{fig:modes}, mode 5). \cite{Kohno2DHub} 
\par
{\it Spin-wave mode of the Mott insulator.--}
The spin-wave mode of the Mott insulator in the 2D Hubbard model can be interpreted as a result of the mode for $\omega>0$ in the LHB (Fig. \ref{fig:modes}, mode i), 
which originates from the upper edge of the spinon-antiholon continuum in the 1D Hubbard model (Fig. \ref{fig:modes}, mode 1), 
and the mode (Fig. \ref{fig:modes}, mode ii) originating from the 1D spinon mode (Fig. \ref{fig:modes}, mode 2). \cite{Kohno2DHub,Kohno1DHub,KohnoSpin} 
\par
Thus, various spectral features of hole-doped and electron-doped systems near the Mott transition can be explained in a unified manner, 
by noting how the spectral weights of the 2D Hubbard model are shifted by next-nearest-neighbor hopping: 
the spectral weights are shifted to higher and lower values of $\omega$ in the momentum regime 
where the values of the Fourier transform of the next-nearest-neighbor hopping integral $t^{\prime}({\bm k})$ are positive and negative, respectively, 
and the shift becomes large where the value of $|t^{\prime}({\bm k})|$ and the spectral weight are large [Eq. (\ref{eq:RPAtd})]. 
The argument can be straightforwardly generalized to the cases where there are further-neighbor hoppings: 
the spectral weights are shifted to higher and lower values of $\omega$ in the momentum regime 
where the values of the Fourier transform of the additional (next-nearest- and further-neighbor) hopping integrals are positive and negative, respectively. 
\par
The key feature to explain the significant asymmetry between hole-doped and electron-doped systems around $\omega=0$ is the behavior of the flat mode around ($\pi,0$) (Fig. \ref{fig:modes}, modes vi, VI, and VI$^{\prime}$). 
Because this mode carries large spectral weights around ($\pi,0$) where the values of the Fourier transform of the next-nearest neighbor hopping integral are negative and large in magnitude [$t^{\prime}({\bm k})\approx -4t^{\prime}<0$], 
its spectral weights are significantly shifted downward by next-nearest-neighbor hopping [Eq. (\ref{eq:RPAtd})] around $\omega=0$. 
Thus, the relative position with respect to the Fermi level is sensitive to the strength of next-nearest-neighbor hopping, which causes remarkable differences between the hole-doped and electron-doped systems around $\omega=0$. 
For the momentum regimes around $(0,0)$ and $(\pi,\pi)$, although the modes around $(0,0)$ and $(\pi,\pi)$ are shifted upward by next-nearest-neighbor hopping 
[Eq. (\ref{eq:RPAtd}), $t^{\prime}({\bm k})\approx 4t^{\prime}>0$; Fig. \ref{fig:modes}, modes I--IV and I$^{\prime}$--IV$^{\prime}$], 
the properties around $\omega=0$ in these momentum regimes are almost unaffected by next-nearest-neighbor hopping, because there is no mode carrying large spectral weights around $\omega=0$ in these momentum regimes. 
\par
As for the Mott transition, the most significant feature is the behavior of the dispersing mode 
whose spectral weights disappear toward the Mott transition (Fig. \ref{fig:modes}, modes 1, i, I, and I$^{\prime}$). \cite{Kohno1DHub,Kohno2DHub,KohnoSpin} 
This characteristic feature, which reflects the spin-charge separation of the Mott insulator, causes \cite{Kohno2DHub} various anomalous features observed in high-$T_c$ cuprates, 
such as the momentum-dependent reduction in spectral weight around $\omega=0$ and the doping-induced states. \cite{DagottoRMP,ShenRMP,FlatbandBi2212,LSCO_FS,ArmitageRMP,ArmitagePRL,Ca2NaCuO2Cl2,XraySWT_PRB,XraySWT_PRL} 
For comparison, in the non-interacting case, where the spectral properties are determined by the single mode carrying the same spectral weight along the dispersion relation regardless of ${\bm k}$, 
neither the band splitting into the UHB and the LHB, nor the momentum-dependent reduction in spectral weight for $\omega\approx 0$, nor the bifurcation of the mode into spinon-like and holon-like branches occurs. 
\par
Although the properties of the ground state might depend strongly on long-range correlations and influence the details of lineshapes and fine structures of the spectral function, 
the overall spectral features which are characterized primarily by the dominant modes (Fig. \ref{fig:modes}, modes 1--5, i--vi, I--VI, and I$^{\prime}$--VI$^{\prime}$) 
should be rather governed by the short-range physics that can be captured by small-cluster calculations. 
Thus, the present results for the overall spectral features would not be significantly affected by whether the ground state is a paramagnetic state such as a Fermi liquid, 
an antiferromagnetically ordered state, or a superconducting state as long as the value of the antiferromagnetic or superconducting order parameter is not so large. 
\par
In this paper, the spectral features of the 2D Hubbard model with next-nearest-neighbor hopping have been simply explained, 
by tracing their origins back to those of the 1D and 2D Hubbard models (Fig. \ref{fig:modes}), 
based on the consideration of how the spectral weights of the 2D Hubbard model are shifted by next-nearest-neighbor hopping. 
Various anomalous features observed in hole-doped and electron-doped high-$T_c$ cuprates \cite{DagottoRMP,ShenRMP,Graf,UniversalFlatbandBi2212,FlatbandBi2212,LSCO_FS,holepocketYBCO,ArmitageRMP,ArmitagePRL,Ca2NaCuO2Cl2,XraySWT_PRB,XraySWT_PRL} 
are naturally interpreted in a unified manner as properties near the Mott transition in a two-dimensional system whose spectral weights are shifted by next-nearest-neighbor hopping. 
\begin{acknowledgments}
This work was supported by KAKENHI (No. 22014015, 23540428, and 26400372) and the World Premier International Research Center Initiative (WPI), MEXT, Japan. 
The numerical calculations were partly performed on the supercomputer at the National Institute for Materials Science. 
\end{acknowledgments}

\begin{references}
\bibitem{NoteForPerturbation} Equation (\ref{eq:RPAtd}) is not the exact perturbation expansion of $G^{-1}({\bm k},\omega)$ up to the first order in $t^{\prime}$, 
because the system does not separate into disconnected parts at $t^{\prime}=0$. Nevertheless, the approximation is useful for the intuitive interpretation of how the next-nearest-neighbor hopping shifts the spectral weights, as described in this paper.
\end{references}

\end{document}